\documentclass[prl,twocolumn,showpacs,superscriptaddress,preprintnumbers,amsmath,amssymb]{revtex4}
\usepackage{graphicx}
\usepackage{dcolumn}
\usepackage{bm}

\def\kt{{k_{_\perp}}}
\def\cO#1{{{\cal{O}}}\left(#1\right)}
\newcommand{\lqcd}{\Lambda_{_{_{\rm QCD}}}}
\newcommand{\dd}{{\rm d} }

\begin{document}
\title{Hadronic single inclusive $k_\perp$ distributions inside one
 jet beyond MLLA}

\author{Fran\c{c}ois Arleo\footnote{On leave from
Laboratoire d'Annecy-le-Vieux de Physique Th\'eorique (LAPTH),
Universit\'e de Savoie, CNRS, B.P. 110,
F-74941 Annecy-le-Vieux Cedex, France}}
\affiliation{CERN, PH department, TH division, CH-1211 Geneva 23}
\author{Redamy P\'erez-Ramos}
\affiliation{Max-Planck-Institut f\"ur Physik, Werner-Heisenberg-Institut,
F\"ohringer Ring 6, D-80805 M\"unchen}
\author{Bruno Machet}
\affiliation{
Laboratoire de Physique Th\'eorique et Hautes \'Energies
\footnote{LPTHE, UMR 7589 du CNRS associ\'ee \`a
l'Universit\'e P.\;et M.\;Curie - Paris\;6 et \`a
l'Universit\'e D.\,Diderot - Paris\;7},
BP 126, 4 place Jussieu, F-75252 Paris Cedex 05
}

\date{December 3rd  2007}

\begin{abstract}
The hadronic $\kt$-spectrum inside one jet is determined including
corrections of relative magnitude $\cO{\sqrt{\alpha_s}}$ with respect
to the Modified Leading Logarithmic Approximation
(MLLA),  at and beyond the limiting spectrum (assuming an infrared
cut-off $Q_0 =\lqcd$ and $Q_0\ne\lqcd$).
The agreement between our results and preliminary measurements by the
CDF collaboration is impressive, much better than at MLLA,
pointing out very small overall non-perturbative contributions.
\end{abstract}


\maketitle

\begin{figure}[h]
\begin{center}
\includegraphics[height=2.5cm]{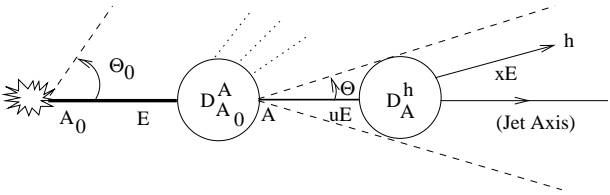}
\caption{\label{fig:distri} Inclusive production of hadron $h$ at angle
$\Theta$ inside a high energy jet of total opening angle
$\Theta_0$ and energy $E$.}
\end{center}
\end{figure}

\begin{figure}
\begin{center}
\includegraphics[height=5cm,width=6.2cm]{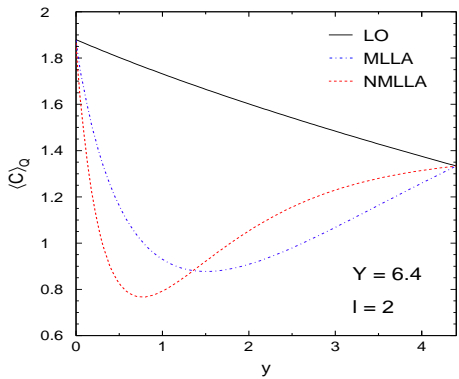}
\caption{\label{fig:CCQ}The color current of a quark jet with $Y_{\Theta_0}
= 6.4$ as a function of $y$ at fixed $\ell=2$.}
\end{center}
\end{figure}

Jet production --~a collimated bunch of hadrons~--
in $e^+e^-$, $e^-p$ and hadronic collisions is an ideal playground
for parton evolution in perturbative QCD (pQCD).
One of the major successes of pQCD is the hump-backed
shape of inclusive spectra, predicted in~\cite{HBP} within MLLA,
and later discovered experimentally~(see e.g.~\cite{KhozeOchs}). Refining the comparison of pQCD calculations
with data taken at LEP, Tevatron and LHC will ultimately allow for
a crucial test of the Local Parton Hadron Duality (LPHD)
hypothesis~\cite{LPHD} and for a better understanding of color
neutralization processes. In this Letter, a class of next-to-next-to-leading
logarithmic (NMLLA) corrections to the single inclusive $\kt$-distribution
of hadrons inside one jet is determined.
Unlike other NMLLA corrections, these terms
better account for recoil effects  and
were shown to drastically affect multiplicities and particle
correlations in jets
\cite{DokKNOCuypersTesima}.
We start by writing the MLLA evolution equations for the fragmentation
function $D_{B}^h\left(x\big/z, zE\Theta_0, Q_0\right)$ of
a parton B (energy $zE$ and transverse momentum $\kt=zE\Theta_0$)
into a gluon (identified as a hadron $h$ with energy $xE$
according to LPHD) inside a jet of energy $E$.
As a consequence of angular ordering in parton cascading,
partonic distributions inside a quark and gluon jet,
$Q,G(z)=x\big/z\ D_{Q,G}^h\left(x\big/z, zE\Theta_0, Q_0\right)$, obey
the system of two coupled  equations \cite{RPR2} (the subscript ${}_y$ denotes
$\partial/\partial{y}$)
\begin{eqnarray}
 Q_{y}&=&\!\!\int_0^1 \dd z\>  \frac{\alpha_s}{\pi} \>\Phi_q^g(z)\>
\bigg[ \Big( Q(1-z)-Q\Big) +  G(z) \bigg],\label{eq:qpr}\\
 G_{y}&=&\!\!\int_0^1 \dd z\> \frac{\alpha_s}{\pi} \>
\bigg[\Phi_g^g(z)(1-z) \Big( G(z) + G(1-z)-
 G\Big)\nonumber\\
&&+n_f\; \Phi_g^q(z)\, \Big(2 Q(z)- G\Big) \bigg],
\label{eq:gpr}
\end{eqnarray}
where $\Phi_A^B(z)$ denote the DGLAP \cite{basics} splitting functions,
$\alpha_s=2\pi\big/ 4N_c\beta_0(\ell+y+\lambda)$ is the one-loop
coupling constant of QCD~\footnote{A 2-loop evaluation
of the splitting functions and $\alpha_s$ would not fit into the present
logic of a systematic expansion in powers of $\sqrt{\alpha_s}$ (see
\cite{APM}).}
 and
\begin{equation*}
\ell=\left(1/x\right)\ ,\ y=\ln\left(\kt\big/Q_0\right)\ ,
\ \lambda=\ln\big(Q_0/\lqcd\big),
\end{equation*}
($Q_0$ being the collinear cut-off parameter), and where
\begin{eqnarray*}
G & \equiv & G(1) = xD_G^h(x,E\Theta_0,Q_0),\cr
Q & \equiv & Q(1) = xD_Q^h(x,E\Theta_0,Q_0).
\end{eqnarray*}
At small $x\ll z$, the fragmentation functions behave as
\begin{equation*}
 B(z) \approx \rho_B^{h} \left(\ln\frac{z}{x},\ln\frac{zE\Theta_0}{Q_0}\right)
=\rho_B^h\left(\ln z + \ell, y\right),
\end{equation*}
$\rho_B^{h}$ being a slowly varying function of two logarithmic
variables $\ln (z/x)$ and $y$ that describes the ``hump-backed''
plateau~\cite{HBP}. In order to better account for recoil effects,
the strategy followed in this Letter is
to perform  Taylor expansions (first advocated for in \cite{Dremin})
of the non-singular parts of the integrands
in~(\ref{eq:qpr},\ref{eq:gpr}) in powers of
$\ln z$ and $\ln(1-z)$, both considered  small  with respect to
$\ell$ in the hard splitting region  $z\sim 1-z =\cO{1}$
\begin{eqnarray}
\label{eq:logic1}
B(z) = B(1) + B_\ell(1) \ln z + \cO{\ln^2z}\ ;\ z\leftrightarrow 1-z\, .
\end{eqnarray}
Each $\ell$-derivative giving an extra  $\sqrt{\alpha_s}$ factor
(see~\cite{RPR2}), the terms $B_\ell(1) \ln z$ and $B_\ell(1)
\ln\left(1-z\right)$ yield NMLLA corrections to the solutions of
(\ref{eq:gpr}).
From (\ref{eq:logic1}) and the expressions of the DGLAP splitting
functions, one gets after some algebra
($\gamma_0^2=2N_c \alpha_s/\pi$)~\cite{APM}
\begin{eqnarray}
Q(\ell,y)&=&\delta(\ell)+\frac{C_F}{N_c}
\int_0^{\ell}\! \dd \ell'\!\int_0^{y}\! \dd y' \gamma_0^2(\ell'+y')
\label{eq:solq}\\
&\times&\Big[
1-\tilde a_1\delta(\ell'-\ell) + \tilde a_2\delta(\ell'-\ell)
\psi_\ell(\ell',y')  \Big]G(\ell',y'),\nonumber\\
G(\ell,y)&=& \delta(\ell)
+\int_0^{\ell}\! \dd \ell'\!\int_0^{y}\! \dd y' \gamma_0^2(\ell'+y')
\label{eq:solg}\\
&\times&\Big[
 1 - a_1\delta(\ell'-\ell) + a_2\delta(\ell'-\ell)\psi_\ell(\ell',y')\Big]
G(\ell',y').\nonumber
\end{eqnarray}
with $\psi_\ell(\ell, y)=G_\ell(\ell, y)/G(\ell, y)$. The MLLA coefficients
${\tilde{a}_1=3/4}$ and $a_1\approx0.935$ are computed in~\cite{RPR2} while
at NMLLA, we get \footnote{Assuming $Q/G=C_F/N_c$. We checked that
$\cO{\sqrt{\alpha_s}}$ and $\cO{\alpha_s}$ corrections affect marginally these coefficients.}:
\begin{eqnarray}
\tilde a_2&=&\frac78+\frac{C_F}{N_c}\left(\frac58-\frac{\pi^2}6
\right)\approx0.42,\\
a_2&=&\frac{67}{36}-\frac{\pi^2}6-\frac{13}{18}
\frac{n_fT_R}{N_c}\frac{C_F}{N_c} \approx 0.06\ .
\label{eq:a2}
\end{eqnarray}
Computing the NMLLA partonic distributions inside a quark and gluon jet,
$Q(z)$ and $G(z)$, is the first step to determine the double differential
spectrum $\dd^2N/\dd{x}\,\dd\Theta$ of a hadron produced  with energy
$xE$ and at angle
$\Theta$ with respect to the jet axis identified with the direction
 of the energy flow (see \cite{APM}).
As shown in~\cite{PerezMachet}, it is given by
\begin{equation}
\frac{\dd^2N}{\dd x\,\dd\ln{\Theta}}=
\frac{\dd}{\dd\ln\Theta}F_{A_0}^{h}\left(x,\Theta,E,\Theta_0\right),
\label{eq:DD}
\end{equation}
where $F_{A_0}^{h}$ is given by the convolution
of two fragmentation functions
\begin{equation}
 F_{A_0}^{h} \equiv
\sum_{A}\int_x^1 \dd u D_{A_0}^A\left(u,E\Theta_0,uE\Theta\right)
 D_A^{h}\left(\frac{x} {u},uE\Theta,Q_0\right),
\label{eq:F}
\end{equation}
$u$ being the energy fraction of the intermediate parton $A$.
$D_{A_0}^A$ describes
the probability to emit $A$ with energy $uE$ off the parton $A_0$
(which initiates the jet),  taking into account the evolution
of the jet between $\Theta_0$ and $\Theta$. $D_{A}^h$ describes the
probability to produce the hadron $h$ off $A$ with energy fraction $x/u$ and
transverse momentum $\kt\approx uE\Theta\geq Q_0$
(see Fig.~\ref{fig:distri}).
As discussed in \cite{PerezMachet}, the convolution
(\ref{eq:F}) is dominated by $u\sim 1$ and therefore
$D_{A_0}^A\left(u,E\Theta_0,uE\Theta\right)$ is given by DGLAP evolution
\cite{basics}. On the contrary, the distribution
$\tilde D_A^h \equiv \frac{x}{u} D_A^{h}\left(\frac{x}{u},uE\Theta,Q_0\right)
=\tilde D_A^h(\ell + \ln u,y)$ at low $x\ll u$ reduces to the hump-backed
plateau,
\begin{equation}
\tilde D_A^h(\ell + \ln u,y) \stackrel{x\ll u}
{\approx} \rho_A^{h}(\ell + \ln u, Y_\Theta + \ln u),
\end{equation}
with $Y_\Theta=\ell+y=\ln E\Theta/Q_0$. Performing the Taylor expansion of
$\tilde D$ to the second order in $(\ln u)$ and plugging it
into Eq.~(\ref{eq:F}) leads to
\begin{eqnarray}
\label{eq:Fdev}
x F_{A_0}^{h}
&\approx& \sum_A\ \int \dd u\, u\, D_{A_0}^A(u,E\Theta_0,uE\Theta)
\tilde D_A^{h}(\ell,y)\\ 
&&\hskip -1cm+\sum_A\ \int \dd u\, u \ln u \, D_{A_0}^A(u,E\Theta_0,uE\Theta)
\frac{\dd \tilde D_A^{h}(\ell,y)}{\dd \ell}\nonumber\\
&&\hskip -1cm+\frac12\sum_A \left[\int \dd u\, u\ln^2 u D_{A_0}^A(u,E\Theta_0,uE\Theta)
\right] \frac{\dd^2 \tilde D_A^{h}(\ell,y)}{\dd\ell^2}.\nonumber
\end{eqnarray}
The first two terms in Eq.~(\ref{eq:Fdev}) correspond
to the MLLA distribution calculated in \cite{PerezMachet}
when  $\tilde D_A^{h}$ is evaluated at NLO and its derivative at LO.
NMLLA corrections arise from their respective calculation at
NNLO and NLO, and, mainly in practice, from  the third line, which is new.
Indeed, since $x/u$ is small, the inclusive spectrum
$\tilde D_A^h(\ell,y)$ is 
the solution of the next-to-MLLA evolution equations (\ref{eq:solq})
and (\ref{eq:solg}). However, because of the smallness
of the coefficient $a_2$ (see (\ref{eq:a2})), $G(\ell,y)$ shows no significant
difference from MLLA to NMLLA. As a consequence, we use the MLLA
expression for $G$.
It is determined here  from a representation in terms of a
single Mellin transform of confluent hypergeometric functions~(see Eq.~(24)
of \cite{finitelambda}), well suited for numerical studies
\footnote{It was also given in~\cite{RPR2} a compact Mellin representation
from which an analytic approximated expression was found using the steepest
descent method~\cite{RPR3}.}.
The NMLLA quark distribution $Q(\ell, y)$ can then be deduced from
$G(\ell, y)$ using (\ref{eq:solq}) and (\ref{eq:solg}), which yields
\begin{eqnarray}\label{eq:ratioqg}
Q(\ell,y)&=&\frac{C_F}{N_c}\left[G(\ell,y)
+\Big(a_1-\tilde a_1\Big)G_\ell(\ell,y)\right.\\ \nonumber
&+&\left.\left
(a_1\Big(a_1-\tilde a_1\Big)+\tilde a_2-a_2\right)G_{\ell\ell}(\ell,y)
\right]+{\cal O}(\gamma_0^2).
\end{eqnarray}
The functions $F_{g}^{h}$ and $F_{q}^{h}$ are related to the gluon
distribution {\it via} the color currents $\langle C\rangle_{g, q}$
defined as:
\begin{equation}
\label{eq:coldef}
x F_{g, q}^{h} = \frac{\langle C\rangle_{g, q}}{N_c}\ G(\ell,y).
\end{equation}
$\langle C\rangle_{g, q}$ can be seen as the average color charge carried
by the parton $A$ due to the DGLAP evolution from $A_0$ to $A$.
Introducing the first and second logarithmic derivatives of
$\tilde D_{A}^{h}$,
\begin{eqnarray*}
\psi_{A,\ell}(\ell,y) &=& \frac{1}{\tilde D_A^h(\ell, y)}
\frac{\dd \tilde D_A(\ell, y)}{\dd\ell}={\cal O}(\sqrt{\alpha_s}),\\
(\psi_{A,\ell}^2+\psi_{A,\ell\ell})(\ell,y)&=&\frac{1}{\tilde D_A^h(\ell, y)}
\frac{\dd^2 \tilde D_A(\ell, y)}{\dd\ell^2}={\cal O}(\alpha_s),
\label{eq:psidef}
\end{eqnarray*}
Eq.~(\ref{eq:Fdev}) can now be written as
\begin{eqnarray}
  x F_{A_0}^{h} &\approx & \sum_A\ 
\Big[ \langle u \rangle_{A_0}^A + \langle u \ln u\rangle_{A_0}^A
\psi_{A,\ell}(\ell,y) \nonumber \\ &+& \frac12 \langle u 
\ln^2 u\rangle_{A_0}^A (\psi_{A,\ell}^2+\psi_{A,\ell\ell})(\ell,y) \Big]
\ \tilde D_{A}^h,
\end{eqnarray}
with the notation
\begin{eqnarray}
  \langle u \ln^i u\rangle_{A_0}^A &\equiv& \int_{0}^{1} \dd u
\ (u\ \ln^i u)\ D_{A_0}^A\left(u,E\Theta_0,uE\Theta\right)\cr
&&\hskip -1cm\approx \int_{0}^{1} \dd u\ (u\ \ln^i u)\ D_{A_0}^A
\left(u,E\Theta_0,E\Theta\right).
\end{eqnarray}
The scaling violation of the DGLAP fragmentation function
neglected in the last approximation is a ${\cal O}(\alpha_s)$
correction to $\langle u\rangle$.
It however never exceeds $5\%$ \cite{APM} of the leading
term and is thus neglected in the following.
Using (\ref{eq:coldef}), the MLLA and NMLLA contributions to the leading
color current of the parton $A_0=g,q$ read
\begin{eqnarray}
&&\hskip -5mm\delta\langle C \rangle_{A_0}^{\rm MLLA-LO} =
N_c\ \langle u \ln u\rangle_{A_0}^g\ \psi_{g, \ell}
 +\ C_{F}\ \langle u \ln u\rangle_{A_0}^q\ \psi_{q, \ell},\cr
&&\delta\langle C \rangle_{A_0}^{\rm NMLLA-MLLA}=
 N_c\ \langle u \ln^2 u\rangle_{A_0}^g
\ (\psi^2_{g, \ell}+\psi_{g,\ell\ell})\cr
 &&\hskip 2cm+\ C_{F}\ \langle u \ln^2 u\rangle_{A_0}^q\ (\psi^2_{q, \ell}
+ \psi_{q, \ell\ell}).
 \label{eq:ccnmlla}
\end{eqnarray}
The MLLA correction, $\cO{\sqrt{\alpha_s}}$,  was determined
in~\cite{PerezMachet} and the NMLLA contribution, $\cO{\alpha_s}$,
to the average color current is new. The latter can be obtained
from the Mellin moments of the DGLAP fragmentation functions
\begin{equation*}
{\cal D}_{A_0}^A(j,\xi)=\int_0^1 \dd u\,u^{j-1} D_{A_0}^A(u,\xi),
\end{equation*}
leading to
\begin{equation}\label{eq:delta2}
\langle u \ln^2 u\rangle_{A_0}^A = \frac{\dd^2}{\dd j^2}{\cal D}_{A_0}^A
(j,\xi(E\Theta_0)-\xi(E\Theta))\bigg|_{j=2}.
\end{equation}
Plugging~(\ref{eq:delta2}) into~(\ref{eq:ccnmlla}), the NMLLA color
currents for gluon and quark jets are determined analytically~\cite{APM}.
For illustrative purposes, the LO, MLLA, and NMLLA average color
current of a quark jet with 
$Y_{\Theta_0}=6.4$ --~corresponding roughly to Tevatron energies~--
is plotted in Fig.~\ref{fig:CCQ} as a function of $y$, at fixed $\ell=2$.
As discussed in~\cite{PerezMachet}, the MLLA corrections to the LO color
current are found to be large and negative.
As expected, the correction $\cO{\alpha_s}$ from MLLA to NMLLA proves
much smaller; it is negative (positive) at small (large) $y$.

\begin{figure}
\begin{center}
\includegraphics[height=5cm,width=6.2cm]{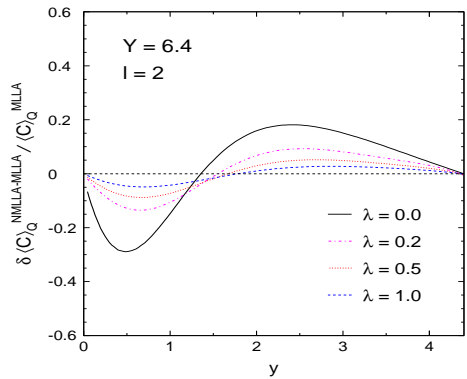}
 \caption{\label{fig:CCQlambda}NMLLA corrections to the color current of
a quark jet with $Y_{\Theta_0}=6.4$ and $\ell=2$ for various
values of $\lambda$.}
\end{center}
\end{figure}

This calculation has also been extended beyond the limiting spectrum,
$\lambda\ne 0$,  to take into account hadronization effects in
the production of ``massive'' hadrons, $m=\cO{Q_0}$~\cite{finitelambda}.
The NMLLA (normalized) corrections to the MLLA result are displayed in
Fig.~\ref{fig:CCQlambda} for different values 
$\lambda=0,0.5,1$. It clearly indicates that
the larger $\lambda$, the smaller the NMLLA corrections.
In particular, they  can be as large as $30\%$ at the limiting
spectrum ($\lambda=0$) but no more than $10\%$ for $\lambda=0.5$.
This is not surprising since $\lambda\ne 0$ ($Q_0\ne\lqcd$) reduces
the parton emission in the infrared sector and, thus, higher-order
corrections.

The double differential spectrum $\dd^2N/\dd{y}\,\dd\ell$,
Eq.~(\ref{eq:DD}), can now be determined from the NMLLA color currents
(\ref{eq:ccnmlla}) using the MLLA quark and gluon distributions
Integrating it over $\ell$ leads to the single inclusive $y$-distribution 
(or $\kt$-distribution) of hadrons inside a quark or a gluon jet:
\begin{equation}
  \left(\frac{\dd N}{\dd y}\right)_{g, q}
= \left(\kt\frac{\dd N}{\dd \kt}\right)_{g, q}
= \int_{\ell_{\rm min}}^{Y_{\Theta_0}-y}\;\dd\ell
\;\left(\frac{\dd^2N}{\dd\ell\, \dd y}\right)_{g, q}.
\label{eq:lmin}
\end{equation}
The MLLA framework does not specify down to which values of
$\ell$ (up to which values of $x$) the double differential
spectrum $\dd^2N/\dd{y}\,\dd\ell$
should be integrated over. Since $\dd^2N/\dd{y}\,\dd\ell$ becomes
negative (non-physical) at small values of $\ell$
(see e.g.~\cite{PerezMachet}), we chose the lower bound $\ell_{\rm min}$
so as to guarantee the positiveness of $\dd^2N/\dd{y}\,\dd\ell$
over the whole $\ell_{\rm min}\le \ell \le Y_{\Theta_0}$ range
(in practice, $\ell_{\rm min}^g\sim 1$ and $\ell_{\rm min}^q\sim 2$). 

\begin{figure}
\begin{center}
\includegraphics[height=9cm,width=8.5cm]{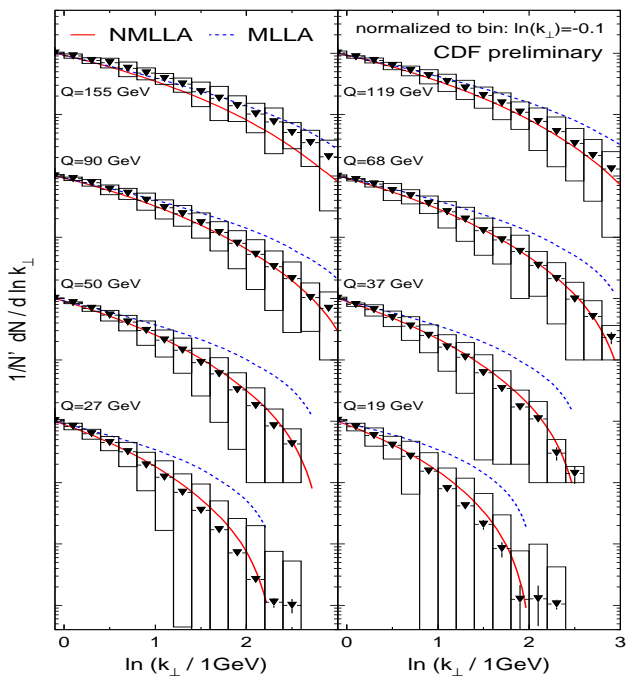}
\caption{\label{fig:CDF-NMLLA}CDF preliminary results for the inclusive
$\kt$ distribution at various hardness $Q$ in comparison to MLLA and
 NMLLA predictions at the limiting spectrum; the boxes are
 the systematic errors (their lower limits at large $\kt$ are cut
for the sake of clarity).}
\end{center}
\end{figure}

Having successfully computed the single $\kt$-spectra including NMLLA
corrections, we now compare the result with existing data.
The CDF collaboration at the Tevatron
recently reported on preliminary measurements  over a wide
range of jet hardness, $Q=E\Theta_0$, in $p\bar{p}$ collisions at
$\sqrt{s}=1.96$~TeV~\cite{CDF}. CDF data, including systematic errors,
 are plotted in
Fig.~\ref{fig:CDF-NMLLA} together with the MLLA predictions of
\cite{PerezMachet} and the present NMLLA calculations, both
at the limiting spectrum ($\lambda=0$) and taking $\lqcd=250$~MeV;
the experimental distributions suffering from large normalization errors,
data and theory are normalized to the same bin, $\ln(\kt/1\,\text{GeV})=-0.1$.
The agreement between the CDF results and the NMLLA distributions over
the whole $\kt$-range is particularly good.
In contrast, the MLLA predictions prove reliable in a much smaller
$\kt$ interval. At fixed jet hardness (and thus $Y_{\Theta_0}$),
NMLLA calculations prove
accordingly trustable in a much larger $x$ interval.

Despite this encouraging agreement with data, the present calculation still
suffers from various theoretical uncertainties, discussed in detail in
\cite{APM}. Among them, the variation of $\lqcd$ --~giving NMLLA
corrections~-- from the default value $\lqcd = 250$ MeV to $150$ MeV and
$400$ MeV affects the normalized $\kt$-distributions by roughly $20\%$
in the largest $\ln(\kt/1~\text{GeV})=3~\text{GeV}$-bin at $Q=100$ GeV. Also, cutting
the integral (\ref{eq:lmin}) at small values of $\ell$ is somewhat arbitrary.
However, we checked that changing $\ell^g_{\text{min}}$ from $1$ to $1.5$ modifies
the NMLLA spectra at large $\kt$ by $\sim 20\%$ only
\footnote{The effect of varying $\ell_{\text{min}}$ is more dramatic at MLLA.}. 
Finally, the $\kt$-distribution is determined with respect to the jet energy
flow from 2-particle correlations (which includes a summation over secondary hadrons), 
while experimentally the jet axis
is determined exclusively from {\it all} particles inside the jet.
The question of the matching of these two definitions at $\cO{\alpha_s}$ accuracy
goes beyond the scope of this Letter.

\begin{figure}
\begin{center}
\includegraphics[height=6.2cm]{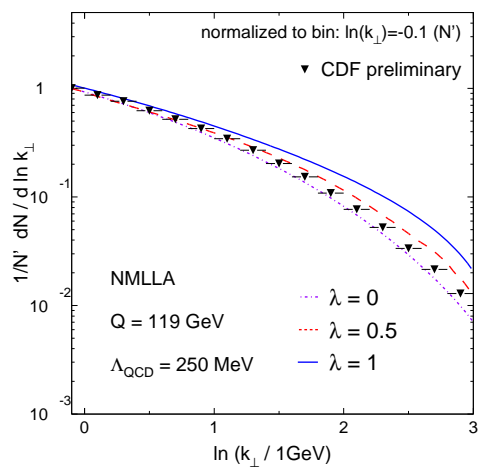}
\caption{\label{fig:CDF-NMLLAlambda}CDF preliminary results ($Q=119$~GeV)
for inclusive $\kt$
distribution compared with NMLLA predictions beyond the limiting spectrum.}
\end{center}
\end{figure}

The NMLLA $\kt$-spectrum has also been calculated beyond the limiting
spectrum, as illustrated in Fig.~\ref{fig:CDF-NMLLAlambda}.
However, 
the best description of CDF preliminary data is reached at the limiting
spectrum,
or at least for small values of $\lambda\lesssim 0.5$, which is not too
surprising since these inclusive measurements 
mostly involve pions. Identifying produced hadrons would offer the
interesting possibility to check a dependence
of the shape of $\kt$-distributions on the hadron species,
such as the one predicted in Fig.~\ref{fig:CDF-NMLLAlambda}.

To summarize, single inclusive $\kt$-spectra inside a jet are
determined including higher-order $\cO{\alpha_s}$ (i.e. NMLLA) corrections
from the Taylor expansion of the MLLA evolution equations and beyond the
limiting spectrum, $\lambda\ne 0$. The agreement between NMLLA predictions
and CDF preliminary data in $p\bar{p}$ collisions at the Tevatron is
very good, indicating very small overall
 non-perturbative corrections. The MLLA evolution equations 
for inclusive enough variables prove once more~(see e.g.~\cite{basics})
to include reliable information
at a higher precision than the one at which they have been deduced.
 \begin{acknowledgments}
\vbox{
{\em Acknowledgments:} We gratefully acknowledge enlightening discussions
with Yu.L.~Dokshitzer, I.M.~Dremin,  S.~Jindariani (CDF),  W.~Ochs and
M.~Rubin.}
\end{acknowledgments}


\end{document}